\documentstyle[aps,prl,graphicx]{revtex}

\begin{document}

\title{ Estimations of electron-positron pair
 production at high-intensity laser interaction with high-$Z$ targets\\}
\author{D.~A.~Gryaznykh, Y.~Z.~Kandiev, V.~A.~Lykov }
\address{Russian Federal Nuclear Center --- VNIITF 
\\ P.~O.~Box 245, Snezhinsk(Chelyabinsk-70), 456770, Russia}
\date{\today}
\maketitle
\begin{abstract}
Electron-positron pairs' generation occuring in the interaction of
\mbox{$10^{18}$--$10^{20}$~W/cm$^2$} laer radiation with high-$Z$
targets are examined.
Computational results are presented for the pair production
and the positron yield from the target with allowance for
the contribution of pair production processes due to electrons and
bremsstrahlung photons.
Monte-Carlo simulations using the {\sc prizma} code confirm the
estimates obtained.
The possible positron yield from high-$Z$ targets irradiated
by picosecond lasers of power \mbox{$10^2$--$10^3$~TW}
is estimated to be \mbox{$10^9$--$10^{11}$}. 
\end{abstract}
\pacs{52.50.Jm, 25.30.Fj, 12.20.Ds}

The possibility of electron-positron pair production by relativistic
electrons accelerated by a laser field has been discussed since
many years~\cite{hora}.It was estimated that the positron
production efficiency can be high~\cite{becker}. The papers cited
considered the case of pair production during oscillations of
electrons in an electromagnetic wave in the focal region of
laser radiation. Here we examine a somewhat different pair
production scenario.

The interaction of high-power laser radiation with matter results
in the production of fast, high-temperature electrons~\cite{wilks}.
Relativistic temperatures of fast electrons $T_f\approx 1\;\text{MeV}$
have been observed in experiments with powerful picosecond
lasers~\cite{wharton}. Self-consistent electric fields confine
these electrons in the target. When the electrons interact with
the matter in a high-$Z$ target, electron-positron pairs are 
produced~\cite{liang}. The annihilation photon spectrum can be used
for diagnostics of the electron-positron plasma.

In the present letter we make estimates of the positron and photon
yield as function of the laser power. We have made an assessment
of the possibility of using high-power \mbox{($10^2$--$10^3$~TW)}
ultrashort-pulse lasers to produce a high-luminosity positron source.
Such sources are required for the production of
slow \mbox{(1--10~eV)} positrons with an intensity of $10^8$
positrons per second. Such positrons find wide applications
for the study of Fermi surfaces, defects and surfaces of various
materials~\cite{puska}.

The interaction of relativistic electrons with matter can lead
to electron-positron pair production in the following two processes:
\begin{displaymath}
\begin{array}{lr}
(\text{i}) &  e^-+Z \rightarrow 2e^- + e^+ +Z ;\\
(\text{ii})&  e^-+Z \rightarrow e+\gamma+Z \rightarrow 2e^- + e^+ + Z.
\end{array}
\end{displaymath}

In Ref.~\cite{dima} analytical and numerical calculations of the total
cross section of the pair electroproduction process are performed
using the differential cross section of Ref.~\cite{bayer}.
According to this work the total cross section of the process (i)
near the threshold equals
\begin{equation}
\sigma_{e\rightarrow 2e e^+} =\frac{7 Z^2 r_e^2 \alpha^2}{2304}
 \frac{\left( E_{0}-2mc^{2} \right)^{3}}{\left( mc^2\right)^{3}},
\label{eq:stot}
\end{equation}
where $r_e$ is the classical electron's radius; $\alpha=1/137$;
$mc^2$ is the electron mass, and $E_0$ is the kinetic energy of the
initial electron. At high energies the cross section grows as~\cite{landau}
\begin{equation}
\sigma_{e\rightarrow 2e e^+}=\frac{28 \pi Z^2 r_e^2 \alpha^2}{27}
\ln^3 E_{0}/ mc^2.
\label{eq:lifchitz}
\end{equation}

The approximation formula
\begin{equation}
\sigma_{e\rightarrow 2e e^+} =5.22 Z^2 \ln^ {3}
 \left( \frac{2.30 + E_0[\text{MeV}]}{3.52} \right)
\: \mu \text{barn}.
\label{eq:approx}
\end{equation}
describes both limits.

\begin{figure}
\begin{center}
\includegraphics[width=8cm,angle=-90]{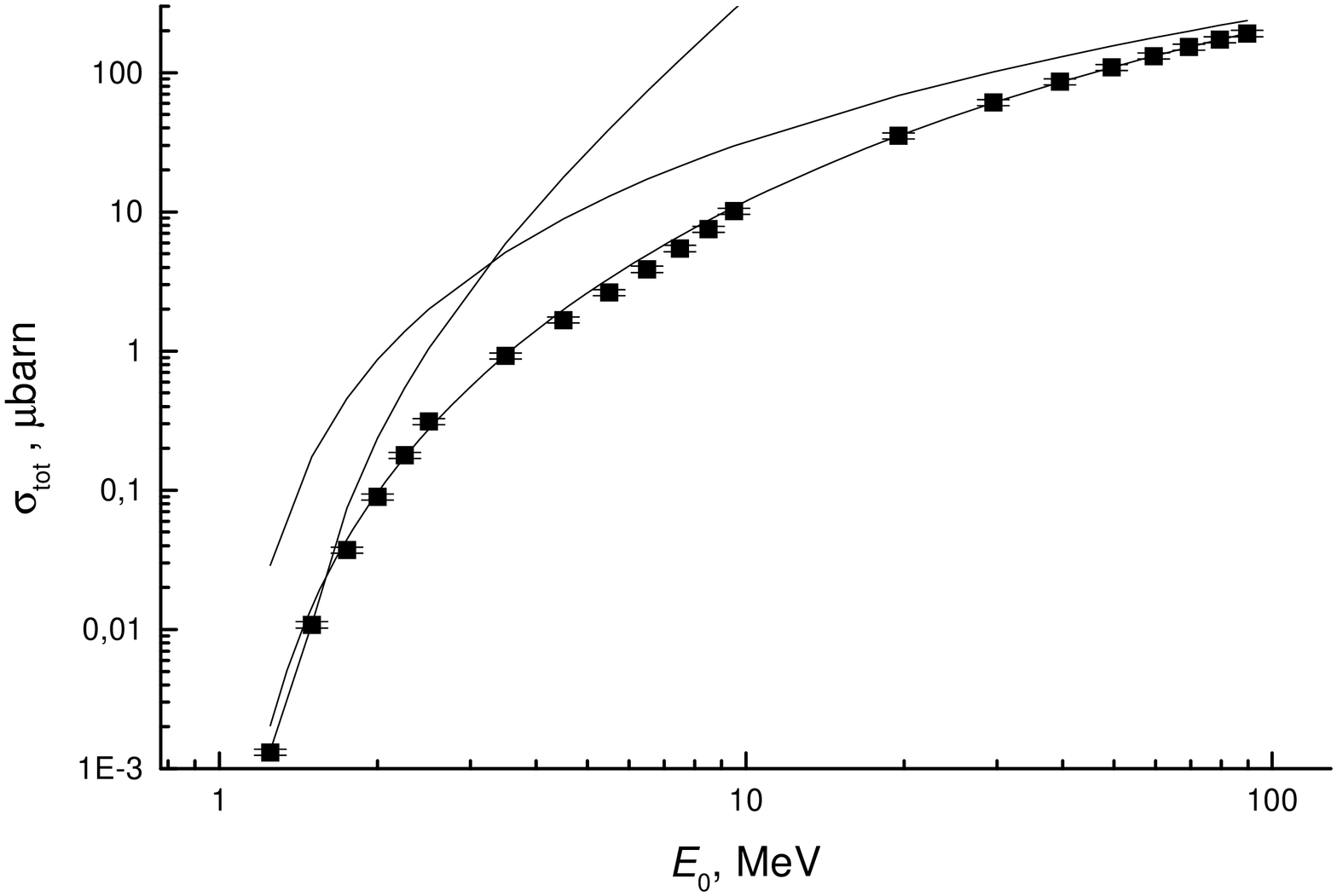}
\caption {Total cross section of electron-positron pairs production
by electron in a Coulomb field of nucleus with $Z=1$;
numerical data, asymptotics and approximation}
\label{fig:stot}
\end{center}
\end{figure}

Fig.~\ref{fig:stot} shows the points obtained by numerically
integrating the exact formulas for the differential section~\cite{dima},
the asymptotic cross sections~(\ref{eq:lifchitz}) and~(\ref{eq:stot}),
and a plot of the approximating function~(\ref{eq:approx}).

The average energy of the positron produced is given by
\begin{equation}
<E_+>=E_0\left(\frac{1}{3}-.0565\ln\frac{E_0}{3mc^2}\right).
\label{eq:emean}
\end{equation}

Let us examine the contribution of the process (i) to the electron-positron
pair production in matter. Let us assume that the fast electrons produced
when the high-intensity laser radiation interacts with matter are confined
by self-consistent electric fields, so that electron moderation
in the target can be treated just as in infinite media.

The probability of pair production during electron moderation in matter
with energy loss from $E_0$ to threshold $2 m c^2$ equals
\begin{equation}
w_e = \int_{2 m c^2}^{E_0} \sigma_{e\rightarrow 2e e^+}
\left(-\frac{dE}{dx} \right)^{-1} \!\!\! n_i \, dE,
\label{eq:intwde}
\end{equation}
where $\sigma_{e\rightarrow 2e e^+}$ is given by~(\ref{eq:approx}),
$n_i$ is the ions density, $dE/dx$ is the electron energy loss
per unit path length.

\begin{figure}
\begin{center}
\includegraphics[width=8cm,angle=-90]{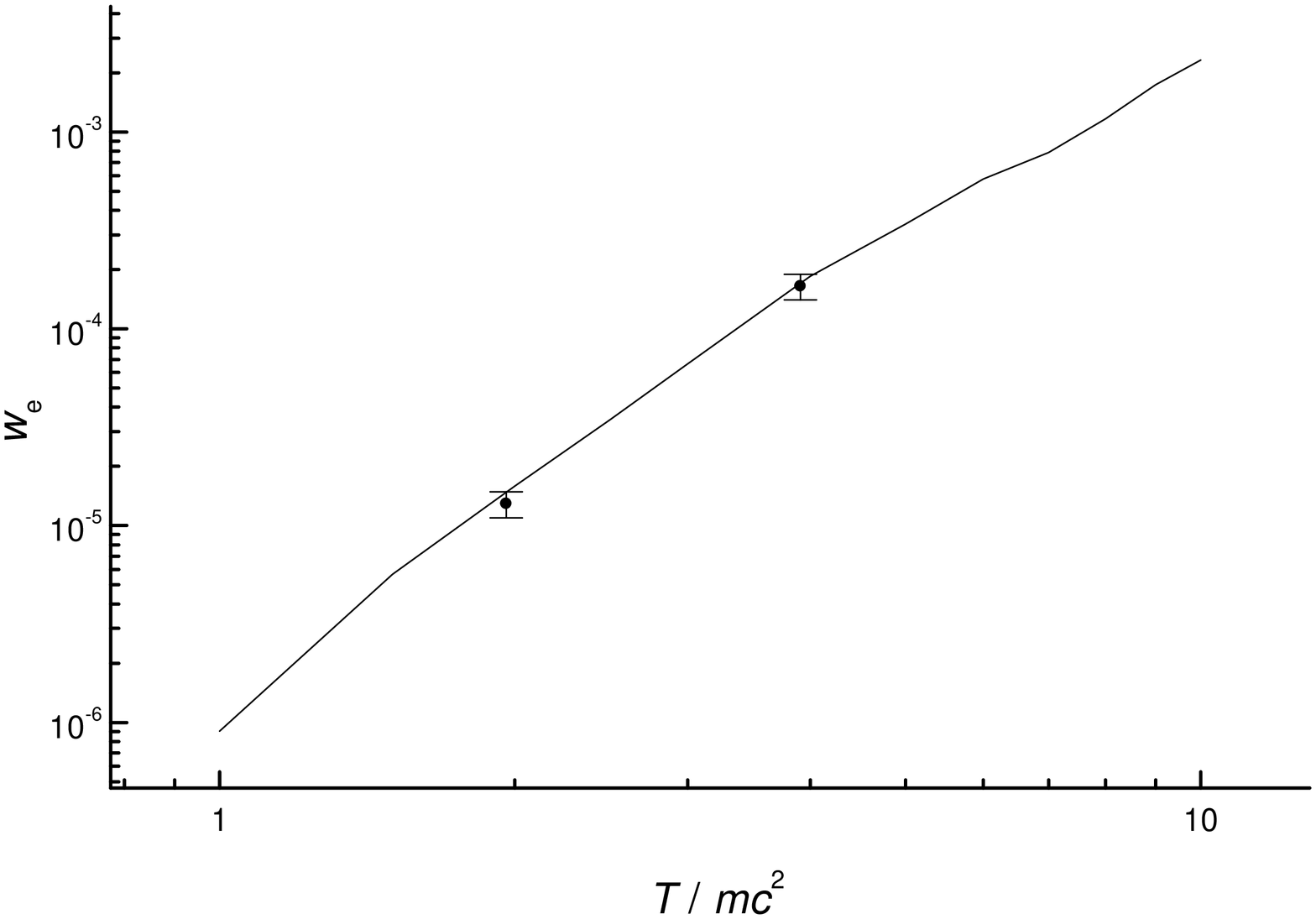}
\caption{Probability of positrons production by electron in the
 Coulomb field of lead's nucleus
versus temperature; points --- {\sc prizma} simulation results}
\label{fig:sf}
\end{center}
\end{figure}

Taking the Rohrlich-Carlsson formula~\cite{akkerman} for $dE/dx$,
we carried out a numerical computation of the integral
in Eq.~(\ref{eq:intwde}) for the case of lead.
Averaging $w_e(E)$ over the relativistic Maxwell distribution
with temperature $T$,
we obtained the number of positrons produced relative
to one initial electron versus temperature.
This dependence is shown in Fig.~\ref{fig:sf}. Performing the same
averaging with weight $<E_{+}>$ from~(\ref{eq:emean}), we obtained the average energy of the
positrons produced.

\begin{figure}
\begin{center}
\includegraphics[width=8cm,angle=-90]{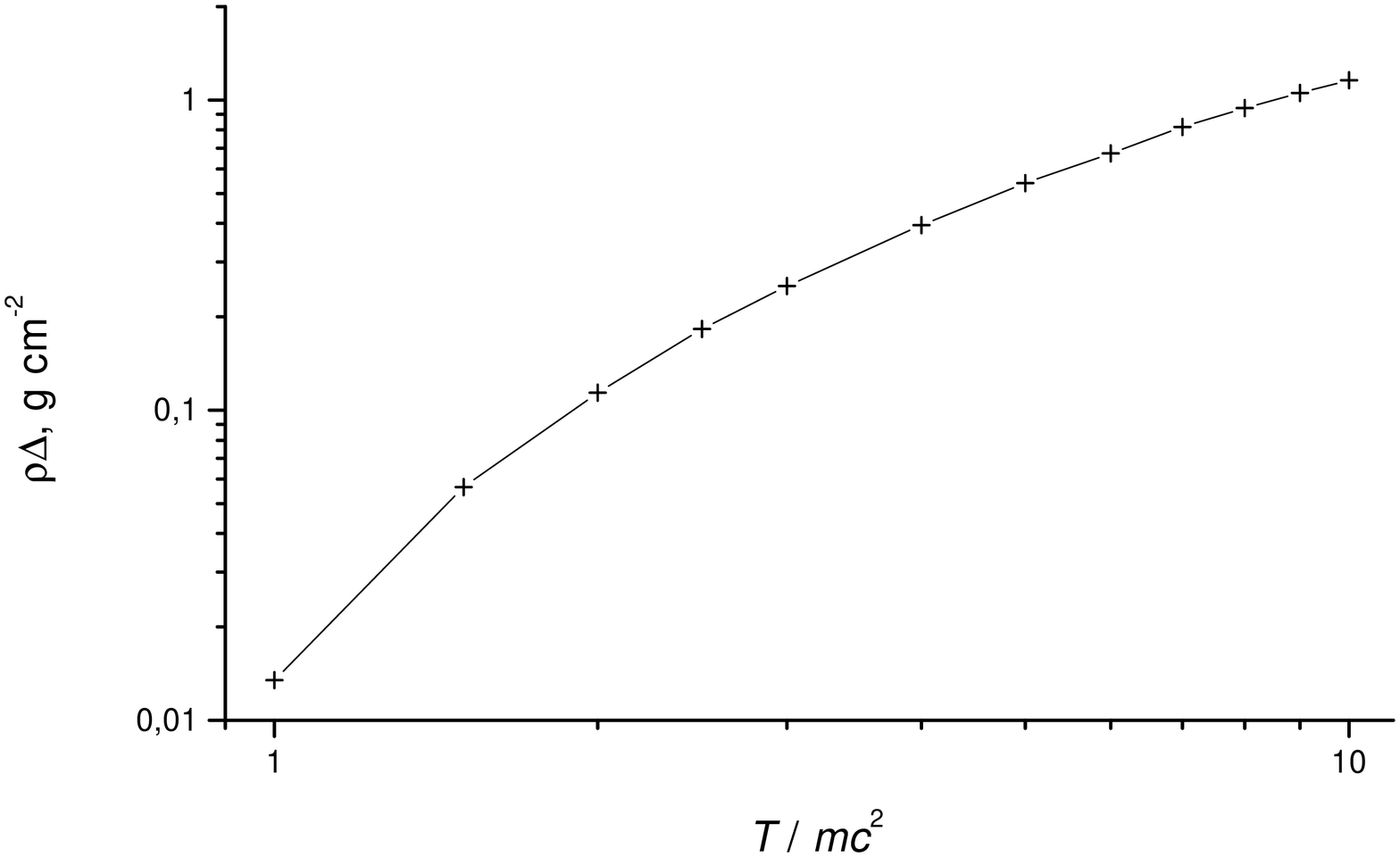}
\caption{The free path length of positrons, produced by electrons,
in lead as a function of temperature}
\label{fig:rhod}
\end{center}
\end{figure}

The average positron energy determines the required thickness of the target,
since the mean free path in matter depends upon energy.
For lead this dependence is determined by~\cite{nemets}
\begin {equation}
\rho\Delta_{e^+}=
\left\{
\begin{array}{lr}
0.412 \left| E \right|^{1.265-0.0954\ln E} & 0.01\le E \le 3, \\
0.53 E - 0.106 & 3 < E < 20,
\end{array}
\right.
\label{eq:rhod}
\end{equation}
where $E$ is given in MeV, $\rho\Delta$ in g~cm$^ {-2} $.
The positron mean free path in lead for different
temperatures of the initial electrons is shown in Fig.~\ref{fig:rhod}.

Let's estimate the probability of pair production by bremsstrahlung photons
(process (ii)). In contrast to the electrons confined in the target by the
self-consistent electric field, photons can escape from it.
The cross section of the process $\gamma\rightarrow e^ + e^-$ is tabulated
in Ref.~\cite{nemets}~(p.~267).
Data on the incoherent photon absorption cross section $\sigma_{a\mathrm{incoh}} $
can also be found there.

The probability of pair production by a single photon of energy
$\varepsilon$ equals
\begin{equation}
w_{\gamma}(\varepsilon)=w_a
\frac{\sigma_{\gamma\rightarrow e^ + e^-}(\varepsilon)}
{\sigma_{a\mathrm{tot}}(\varepsilon)},
\label{eq:wphots}
\end{equation}
where 
$\sigma_{a\mathrm{tot}}(\varepsilon)
=\sigma_{\gamma\rightarrow e^ + e^-}(\varepsilon)+
\sigma_{a\mathrm{incoh}}(\varepsilon)$,
$w_a(\varepsilon)=1-\exp (-\sigma_{a\mathrm{tot}}
(\varepsilon) n_i \Delta)$,
$\Delta$ is the thickness of the target.
For an infinite target
\begin{equation}
 w_{\gamma}^{\infty} =\frac{\sigma_{\gamma\rightarrow e^ + e^-}}
 {\sigma_{a\mathrm{tot}}}.
\label{eq:pbinf}
\end{equation}

\begin{figure} \begin{center}
\includegraphics[width=8cm,angle=-90]{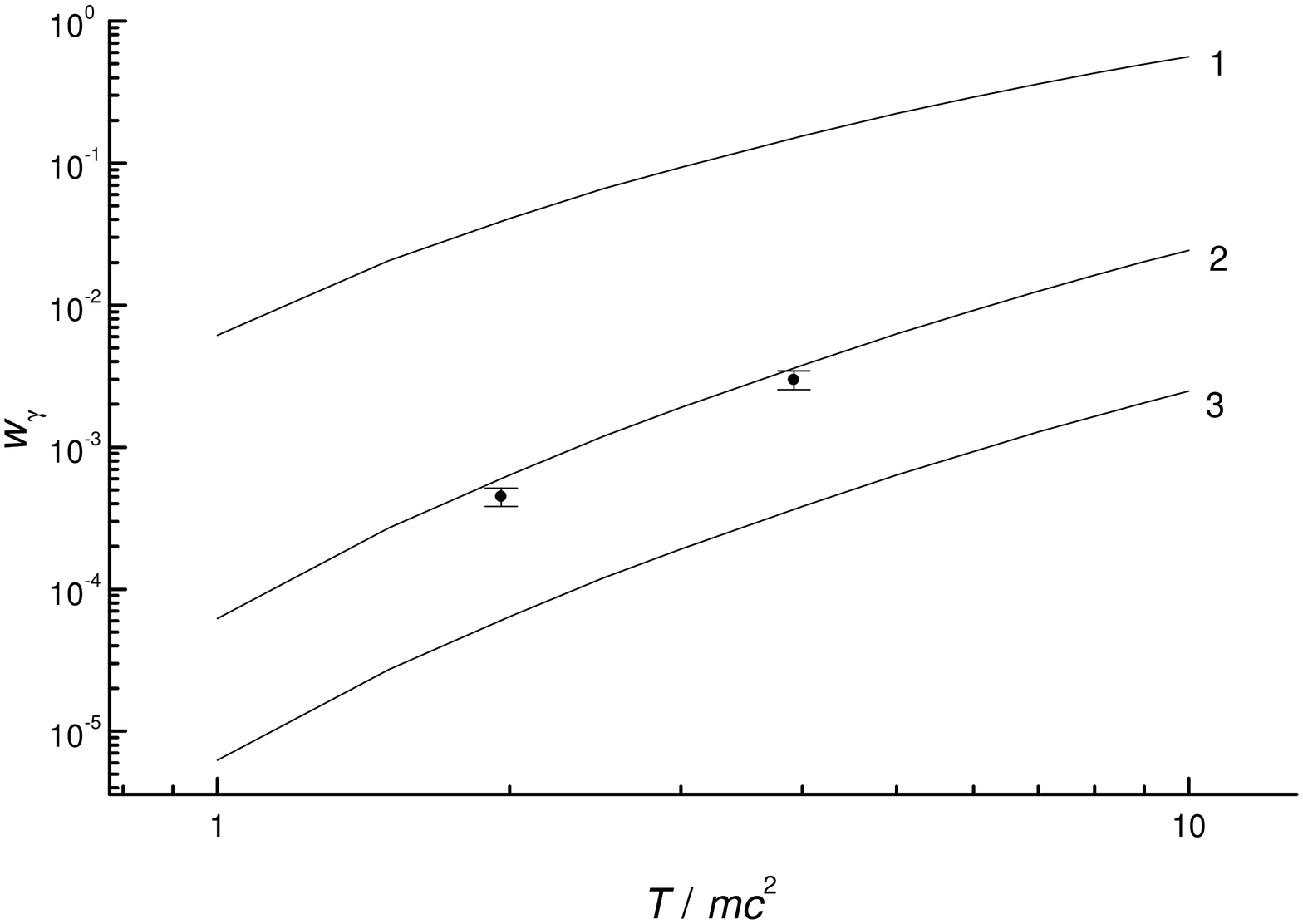}
\caption{Probability of positrons production by electron
through bremsstrahlung photon in lead
for thicknesses $\rho\Delta=\infty$(curve~1);
$\rho\Delta=$3.~g~cm$^{-2}$(curve~2); $\rho\Delta=$0.3~g~cm$^{-2}$(curve~3) 
versus temperature; points --- {\sc prizma} simulations with sphere target
$\rho R=$2.2~g~cm$^{-2}$}
\label{fig:wphot}
\end{center} \end{figure}

We take the photon spectrum in the form
\begin{equation}
dN/d\varepsilon \simeq \epsilon_{e\rightarrow e\gamma} T^{-1}
 \exp (-\varepsilon/T),
\label{eq:phspect}
\end{equation}
where 
$\epsilon_{e\rightarrow e\gamma}=3\times10^{-4}ZT/mc^2$
is the ratio of the total bremsstrahlung power radiated
to the total power in the incident
electron beam determined in Sec.~\mbox{(IV-20)} of Ref.~\cite{koch}.
Averaging $w_{\gamma}(\Delta,\varepsilon)$ over spectrum~(\ref{eq:phspect}),
we obtain
\begin{equation}
w_{\gamma}(\Delta,T) \simeq \epsilon_{e\rightarrow e\gamma}
T^{-1} \int_{2 m c^2}^{+ \infty}\exp(-\varepsilon/T)
w_{\gamma}(\Delta,\varepsilon) d\varepsilon.
\label{eq:phmax}
\end{equation}
The dependence of the number of positrons produced by bremsstrahlung photons
relative to one initial electron upon temperature for an infinite slab
and two thicknesses is presented in Fig.~\ref{fig:wphot}.

The results of the estimation of the number of positrons produced
can be used to estimate the number of annihilation photons
in targets with thickness greater than the positrons mean free path 
(see Fig.~\ref{fig:rhod})
but less than that of photons ($\approx 6$~g~cm$^{-2}$ for lead).
The channel (ii) predominates here. For thickness about \mbox{2--3 g cm$^{-2}$}
the photon yield reaches 0.04\% per one source electron 
with temperature~$T\simeq$~1~MeV.

To check the estimates, calculations were performed using 
the {\sc prizma} code~\cite{kandiev},
which simulates all basic electron, photon and positron transport
and production processes for any geometry (one-, two- and three-dimensional)
by the Monte Carlo method.
The calculation were perfomed for a lead sphere with radius $R=0.2$~cm and
 an electron source with temperature $T=1$ and $T=2$~MeV at the center.
The results are presented
in Figs.~\ref{fig:sf},\ref{fig:wphot}. They are in good agreement
with our estimates.

According to Ref.~\cite{wilks}, the temperature of fast electrons arising
during interaction of laser radiation with matter is about
\begin{equation}
T_f \simeq mc^2 \big[ (1+0.7q_{18})^{\frac{1}{2}}-1\big],
\end{equation}
where $q_{18}$ is the laser power density in \mbox{$10^{18}$ W cm$^{-2}$}.
When a laser pulse with energy~$E_l$[J], duration~$\tau$[psec] is focused into
a circle of diameter $d_f[\mu$m], the intensity equals
$q_{18}=400\:E_l/\pi d_f^2 \tau$.
The number of electrons produced is determined by
\mbox{$N_e=A_f E_l/\langle E_f\rangle$},
where $A_f$ is the efficiency of laser energy conversion to fast electrons,
$\langle E_f \rangle$ is the average energy of fast electrons.

As a target we propose a sphere with conical cavity into which
laser radiation is focused~\cite{lykov}.
Such a target gives $A_f\approx 0.3$, high luminosity
and isotropic yield of photons and positrons from the surface.
 The target must have high $Z$,
and its optimal diameter is determined by experiment tasks and laser power.

\begin{figure} \begin{center}
\includegraphics[width=8cm,angle=-90]{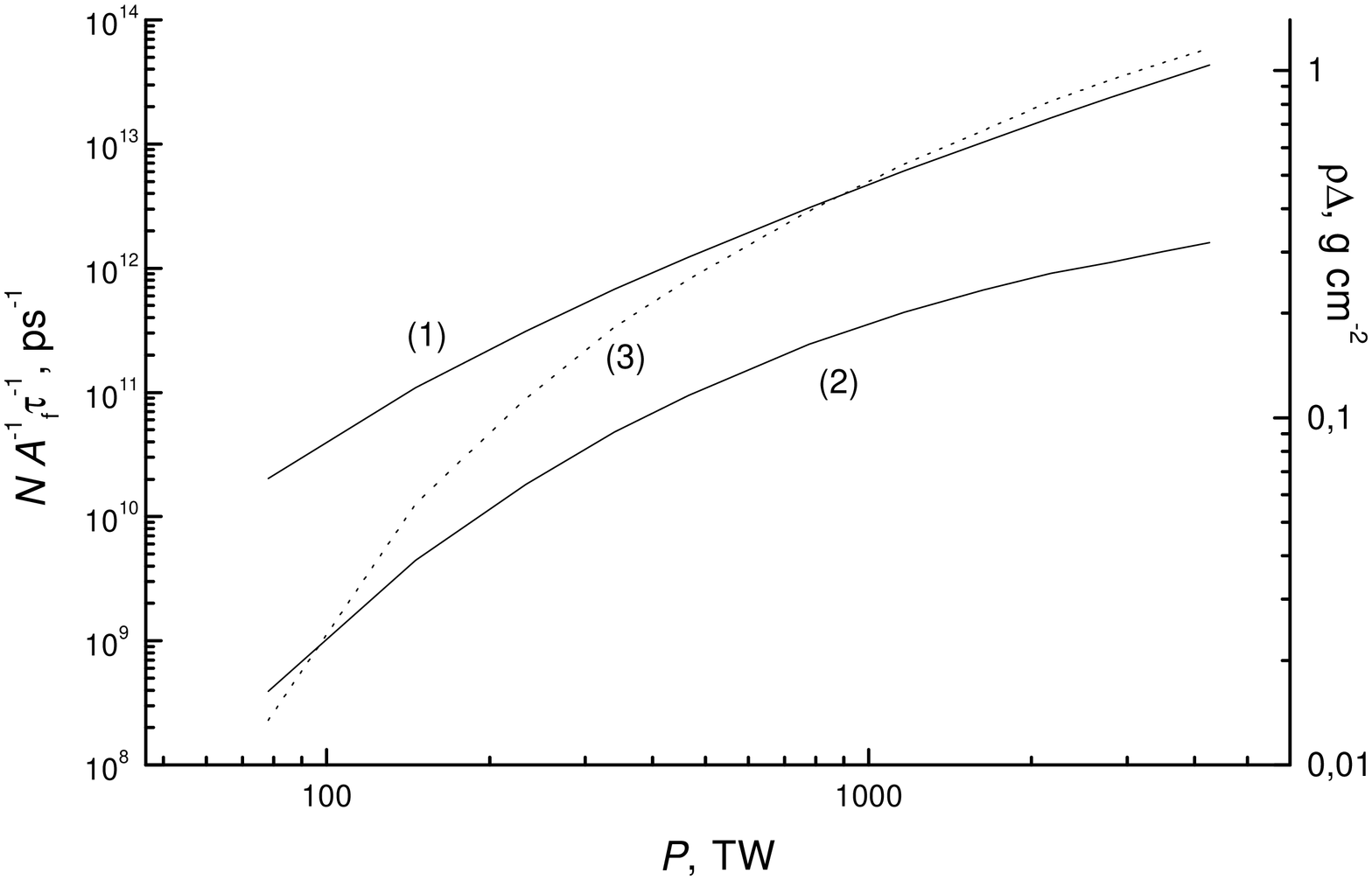}
\caption{Dependence of photon (1) and positron (2) yield
 $N/A_f \tau$ on laser power. Curve (3) shows the optimum size of a target
 for a positron source.}
\label{fig:lasph}
\end{center} \end{figure}

To detect annihilation photons the target has to be of size
\mbox{$\rho R\approx 2$--3 g cm$^{-2}$}.
The dependence of annihilation photon yield $N_{\gamma}$
 divided by $A_f$ and $\tau$,
on laser power is shown in Fig.~\ref{fig:lasph}.
The focal spot is $d_f=30\;\mu$m.
The photon yield is about of \mbox{$10^{10}$--$10^{12}$}
for a \mbox{power $10^2$--$10^3$ TW} picosecond laser.

The positron yield from the target can be estimated as
\begin{equation}
N_+ \simeq N_e \frac{\rho\Delta_{e^+}}{\rho\Delta_{e^+}+\rho\Delta}
\left( w_e+w_{\gamma}^{\infty}\frac{\rho\Delta}{\rho\Delta_{\gamma}}
 \right).
\label{eq:outp}
\end{equation}
Here $\Delta_{e^+,\gamma}$ are the positron and bremsstrahlung
photon mean free paths.
The target for positron production must be of order $\Delta_{e^+}$
in size (see Fig.~\ref{fig:rhod}).
The dependence of positron yield $N_{+}$ divided by $A_f$ and $\tau$,
on laser power is shown in Fig.~\ref{fig:lasph}.
The dotted line in this figure shows the optimal target size $\rho\Delta$
for such an experiment.
The positron yield reaches \mbox{$10^{9}$--$10^{11}$} for
\mbox{powers $10^2$--$10^3$ TW} picosecond laser.

Since the target could be smaller than existing positron sources,
the laser positron source can have a very high brightness.
The efficiency of conversion of fast positrons (MeV) to
slow \mbox{(1--10~eV)} can be as high as $10^{-2}$~\cite{schultz}. 
Therefore, to produce a quasi steady-state source of
slow positrons  with an intensity of $10^8$ positrons per second requires
a laser with energy of \mbox{10--30 J} in \mbox{10--30 fs} pulse
with a repetition frequency \mbox{10-30 Hz}. Undoubtedly, such a source
would be usefull for fundamental and applied researches in
 solid state physics, electronics, chemistry and biology.

\acknowledgements

One of the authors (VL) is grateful to Prof.~H.~Hora for a helpful discussion
and his interest to the work.
This work was supported by the International Science and Technology Center, Project \#107-94.

\end{document}